\documentclass[prl,twocolumn,aps,amssymb,footinbib,showpacs]{revtex4}
\usepackage{amssymb}
\usepackage{graphicx}
\usepackage{amsmath}
\usepackage{times}
\usepackage{color}
\usepackage{subfigure}
\usepackage{setspace}
\usepackage{bm}
 
\begin{document}

\newcommand {\beq} {\begin{equation}}
\newcommand {\eeq} {\end{equation}}
\newcommand {\bqa} {\begin{eqnarray}}
\newcommand {\eqa} {\end{eqnarray}}
\newcommand {\da} {\ensuremath{d^\dagger}}
\newcommand {\ha} {\ensuremath{h^\dagger}}
\newcommand {\adag} {\ensuremath{a^\dagger}}
\newcommand {\no} {\nonumber}
\newcommand {\ep} {\ensuremath{\epsilon}}
\newcommand {\ca} {\ensuremath{c^\dagger}}
\newcommand {\ga} {\ensuremath{\gamma^\dagger}}
\newcommand {\gm} {\ensuremath{\gamma}}
\newcommand {\up} {\ensuremath{\uparrow}}
\newcommand {\dn} {\ensuremath{\downarrow}}
\newcommand {\ms} {\medskip}
\newcommand {\bs} {\bigskip}
\newcommand{\kk} {\ensuremath{{\bf k}}}
\newcommand{\rr} {\ensuremath{{\bf r}}}
\newcommand{\kp} {\ensuremath{{\bf k'}}}
\newcommand {\qq} {\ensuremath{{\bf q}}}
\newcommand{\nbr} {\ensuremath{\langle ij \rangle}}
\newcommand{\ncap} {\ensuremath{\hat{n}}}
    
\begin{abstract}
  We theoretically study the many-body effects of electron electron
  interaction on the single particle spectral function of doped
  bilayer graphene. Using random phase approximation, we calculate the
  real and imaginary part of the self-energy and hence the spectral
  function. The spectral function near the Fermi surface shows the
  usual quasiparticle peak, establishing doped bilayer graphene, in
  contrast to the unstable neutral system, to be a Fermi liquid. Away
  from the Fermi surface, an additional broad plasmaron peak is
  visible in the spectral function. From the low energy behaviour of
  the self-energy we calculate the quasiparticle residue and the
  effective mass of the quasiparticles as a function of carrier
  density. We present results for both the on-shell and the off-shell
  approximation for the quasiparticle renormalization.

\end{abstract}

\title{Correlations, Plasmarons, and  Quantum Spectral Function in Bilayer Graphene }
\date{\today}
\author{  Rajdeep Sensarma, E. H. Hwang, and S. Das Sarma}  
 \affiliation{ Condensed Matter Theory Center, Department of Physics, University of Maryland, College Park, USA 20742}
\maketitle

\section{}

Bilayer graphene (BLG) has attracted both experimental and theoretical
interest as a novel 2D gapless chiral electron-hole system with
parabolic dispersion~\cite{review}, where carrier density can be
easily tuned by applying a gate voltage. From a many-body interaction
physics perspective, BLG is particularly intriguing as it lies between
the chiral and gapless single layer graphene (SLG), which has
linear band dispersion, and 2D semiconductor-based electron gas (2DEG)
systems, which are non-chiral and gapped, but typically have quadratic
band dispersion~\cite{ando:rmp}. In addition, the interaction
parameter in BLG can be simply tuned by changing the carrier density whereas in SLG the
interaction parameter is the fine structure constant which is fixed
for a given substrate.  BLG is therefore the ideal system to
study chiral interaction physics since interaction effects in SLG are
typically small~\cite{review}. It is therefore quite surprising that
very little BLG interaction physics has so far been studied in the
literature, perhaps because of serious technical difficulties of
treating both intraband and interband virtual processes on an equal
footing in a gapless system.  In this Letter we provide the first
theoretical treatment of the many-body interaction effects on the BLG
single-particle properties as a function of carrier density.  Our
work, being directly comparable to experimental data, should motivate
experimental studies of BLG many-body properties.

An important many-body property of BLG is the single particle spectral
function $A(\kk,\omega)$, which measures the probability of finding an
electron (hole) with a momentum $\kk$ and energy $\omega$ in the
system (We take $\hbar=1$ throughout.) The spectral function, which gives detailed information about
electronic structure and many-body renormalization, can be measured by
angle resolved photo-emission spectroscopy (ARPES)~\cite{ARPESref}.
In the absence of interaction, the spectral function is simply a delta
function giving the non-interacting band energy at momentum $\kk$.
The modification of the spectral function due to electron-electron
interactions is a matter of great interest both
theoretically and experimentally.  A pressing question in this
respect is whether BLG behaves like a Fermi liquid, where
phase space restrictions near the Fermi surface lead to sharply
defined quasiparticles with a finite spectral weight.
 
In this letter, we calculate the carrier density dependent single
particle spectral function of doped BLG, taking into account
electron-electron interactions within the dynamical random phase
approximation (RPA). We find that the BLG spectral function has a
broad plasmaron peak, i.e. a peak associated with the quasiparticle
carrying a cloud of virtual plasmons, in addition to the Landau
quasiparticle peak. We obtain the quasiparticle renormalization and
the effective mass, which completely define the low energy coherent
part of the spectral function around the Fermi surface, and study
their variation with the carrier density. We find that quasiparticle
spectral weight decreases with decreasing density while the effective
mass shows a non-monotonic behaviour. The effective mass remains
smaller than the bare mass down to relatively low densities with a
maximum renormalization of $\sim 20\%$, thus showing that the system
remains a weakly interacting Fermi liquid even at low densities.
Earlier BLG many-body theoretical work restricted
itself~\cite{effmass} to either Hartree-Fock or static screening
theories neglecting dynamical correlations, which is known to be
inadequate in interacting electron systems where dynamical screening
is important.

A key question for BLG, indeed for any electronic material, is whether
the presence of interaction preserves the Fermi liquid behavior, where the system manifests one-to-one correspondence with the noninteracting system in its
low-energy properties, or leads to an exotic non Fermi liquid.
This question has recently
been addressed~\cite{nfl} in the literature for neutral undoped BLG
which exists only at the Dirac point.  Not surprisingly such a pure
zero-density Fermi system is found to be a non-Fermi-liquid in the
presence of interaction.  We show in the current work that this Dirac
point non-Fermi liquid behavior of neutral BLG is unstable to the
existence of any finite doping, and doped BLG is always a Fermi liquid
in the presence of any finite carrier density (which is always true in
any physical BLG system).  Our finding that the BLG is a Fermi liquid
similar to regular non-chiral 2DEG systems is directly experimentally
verifiable through ARPES since the singularity structure of the
theoretical spectral function is very different in the doped and the
undoped situations.  We find that the neutral non-Fermi liquid fixed
point is an unstable set of measure zero, which will always become a
Fermi liquid fixed point in the presence of density fluctuations.
This ia an important new result since density fluctuations are
invariably present in all real BLG samples due to thermal and disorder
effects.
 


We work with the low energy two band parabolic approximation to the
BLG Hamiltonian where the dispersions of the bands are given by
$E_{s\kk}=sk^2/2m_0$ where $s=\pm 1$ denotes the conduction and
valence bands and $m_0$ is the non-interacting BLG mass with
$m_0\simeq 0.033 m_e$, $m_e$ being the mass of a free electron. The
chiral band wave functions are given by
$\psi^\dagger_{s\kk}=(e^{i2\theta_\kk},s)/\sqrt{2}$, where
$\theta=\tan^{-1}(k_y/k_x)$ is the azimuthal angle in the momentum
space. Each band has a degeneracy factor of $g=4$ ($2$ for spin and $2$
for valley). We note that the 2-band parabolic BLG approximation is well-justified because we are interested in the low-energy behavior, and at high doping densities, where the parabolic approximation fails, the BLG dispersion becomes linear with the system behaving like the SLG whose many-body properties have already been studied in the literature~\cite{review,slgspectral}.

\begin{figure}[t!]
\includegraphics[width=0.23\textwidth]{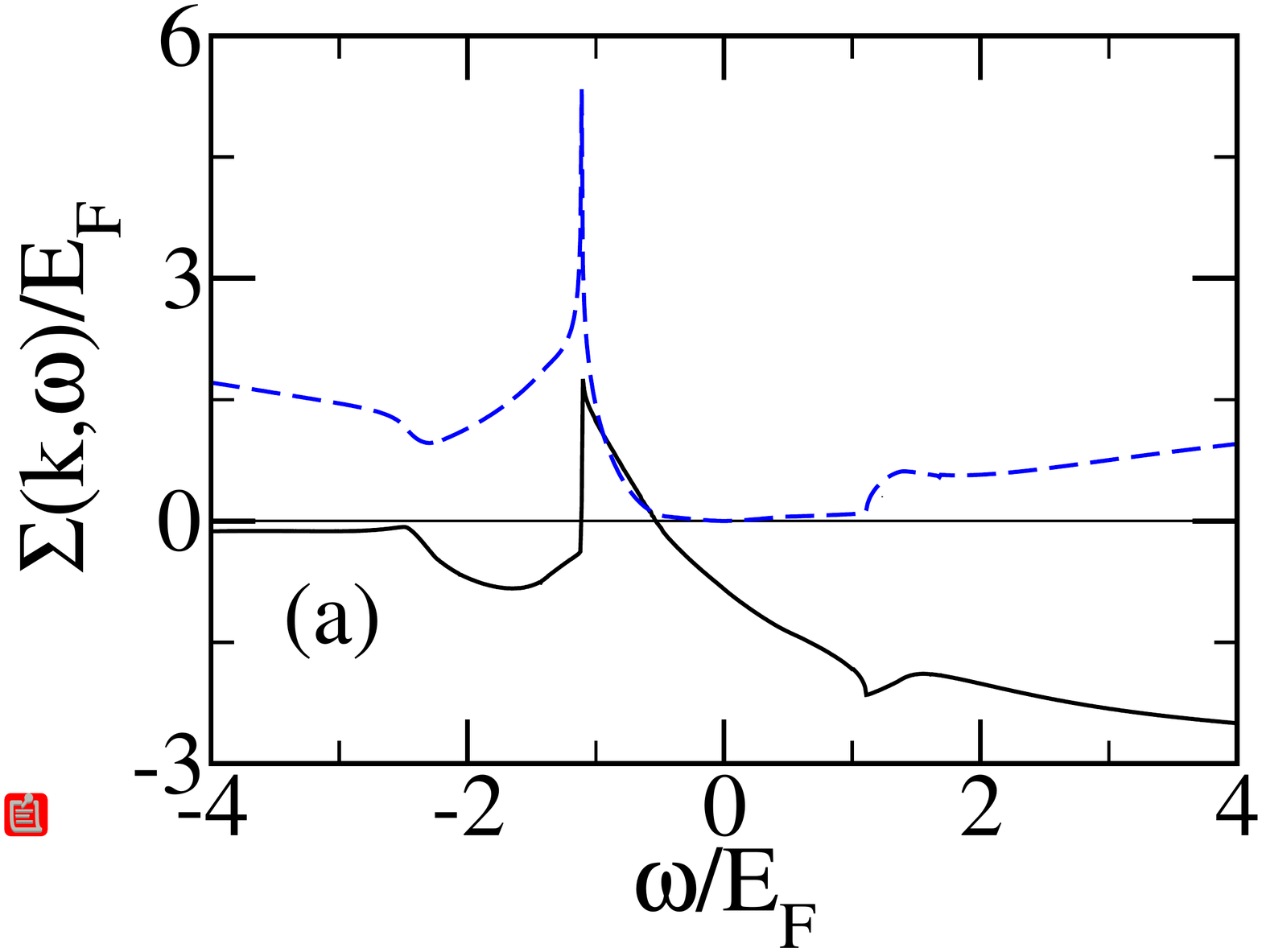}
\includegraphics[width=0.23\textwidth]{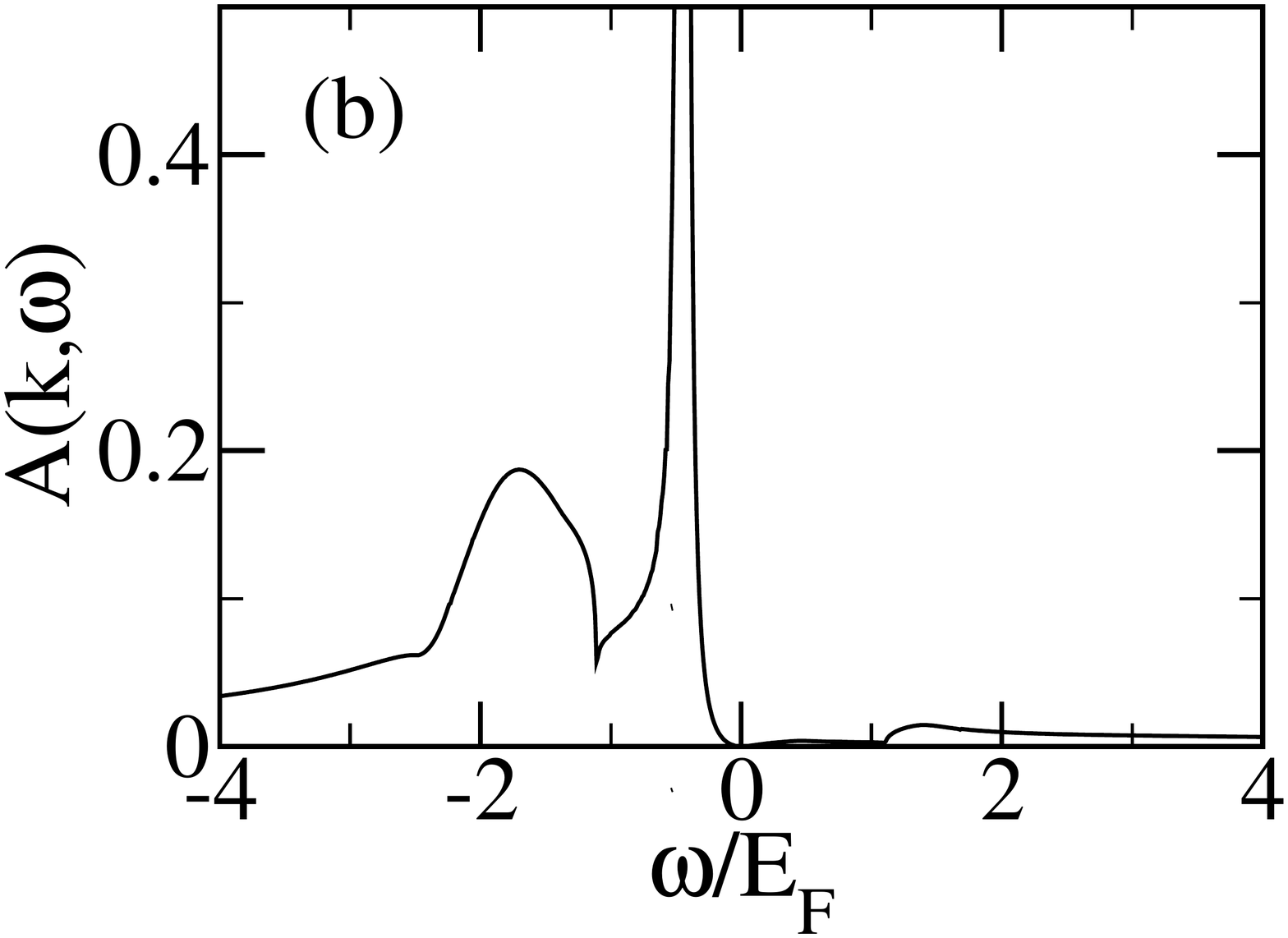}
\includegraphics[width=0.23\textwidth]{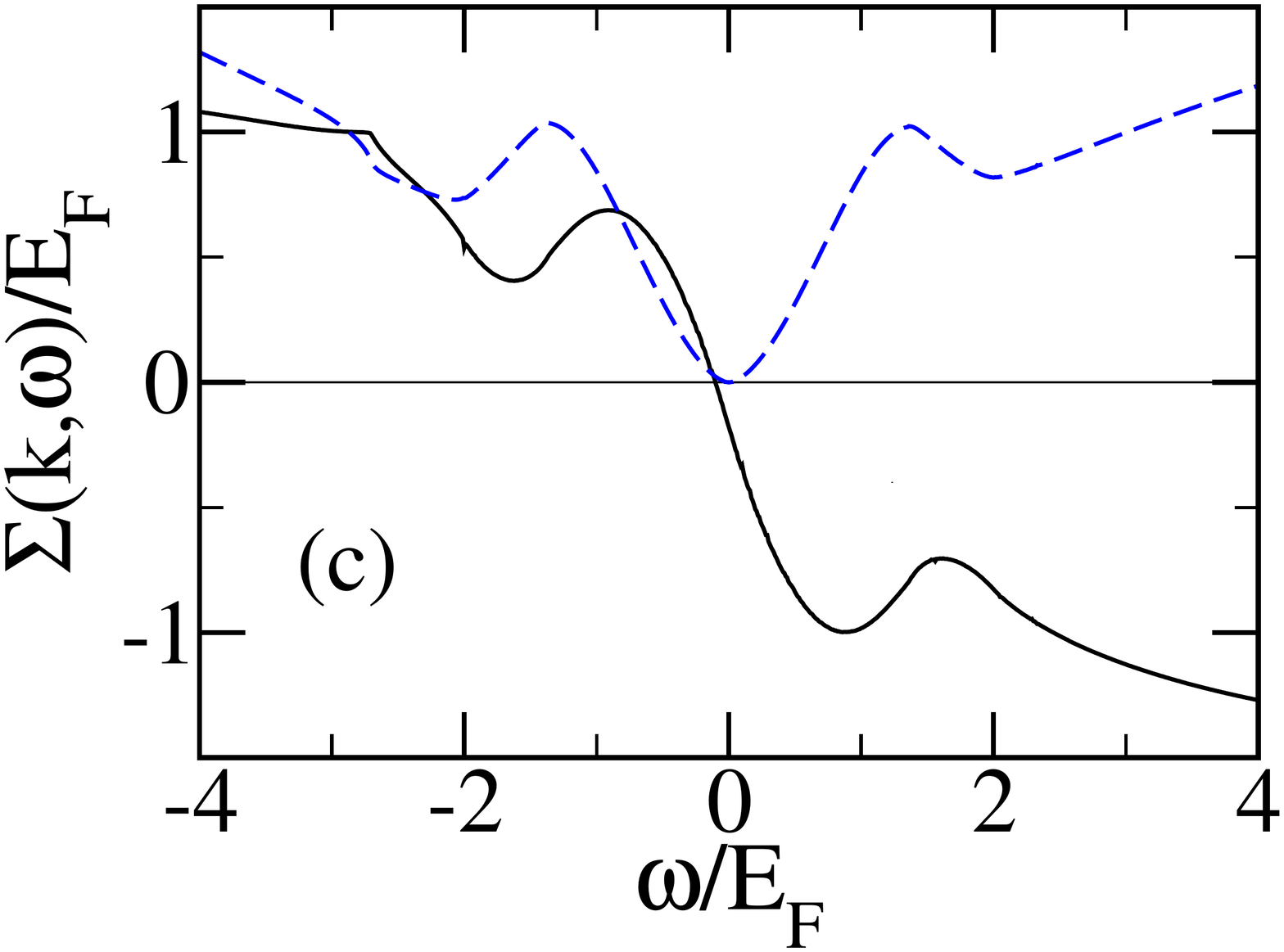}
\includegraphics[width=0.23\textwidth]{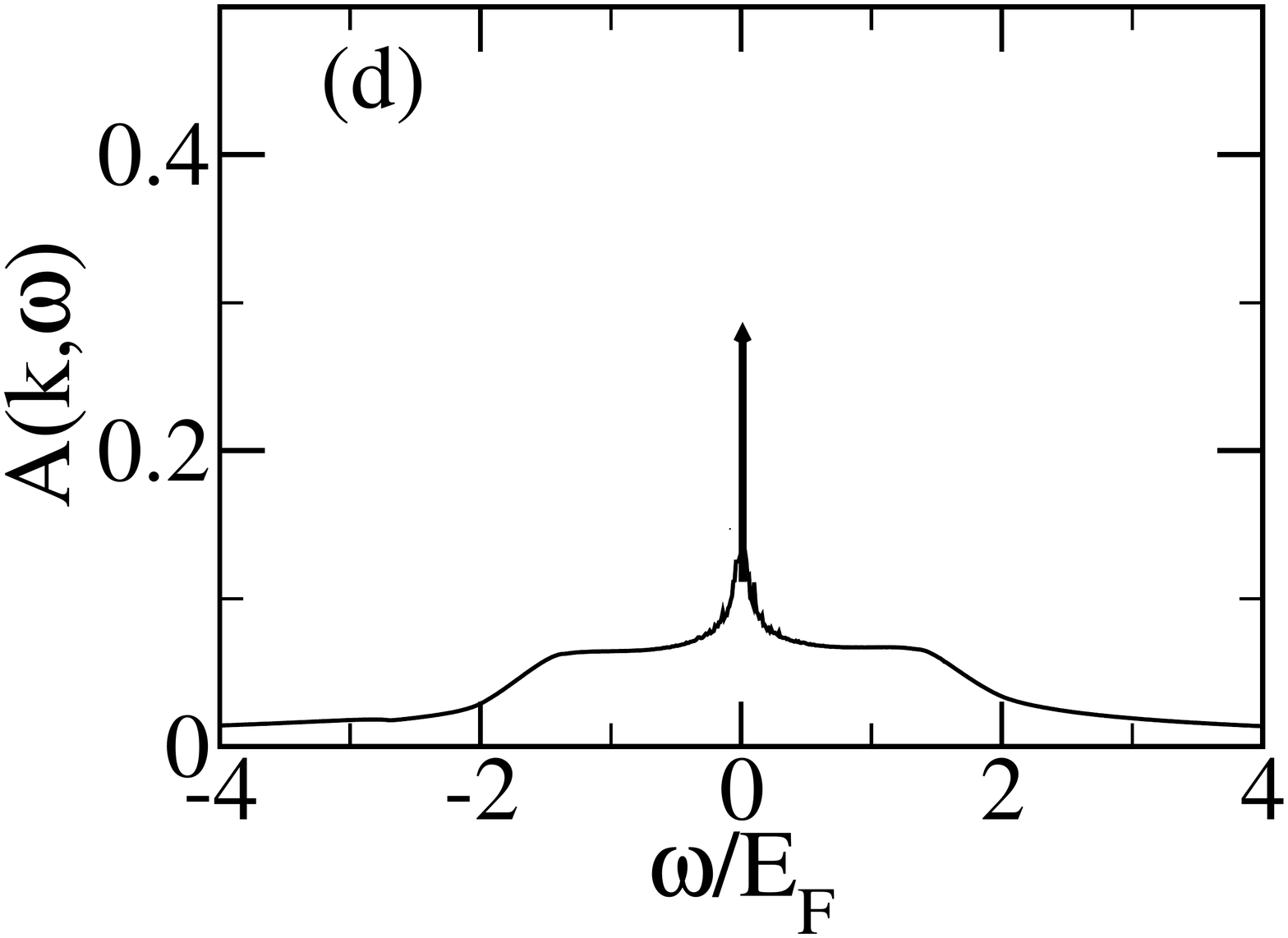}
\includegraphics[width=0.23\textwidth]{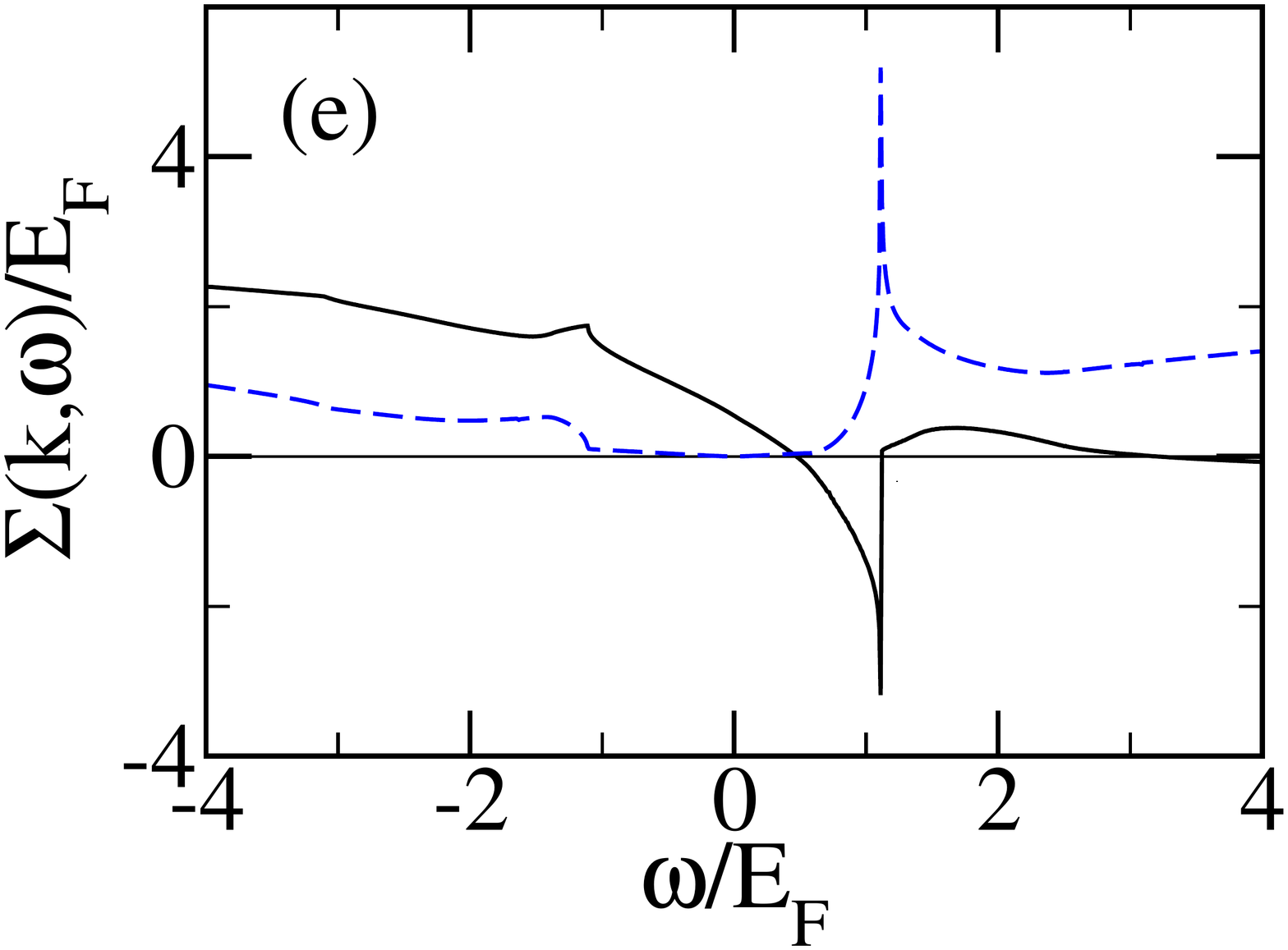}
\includegraphics[width=0.23\textwidth]{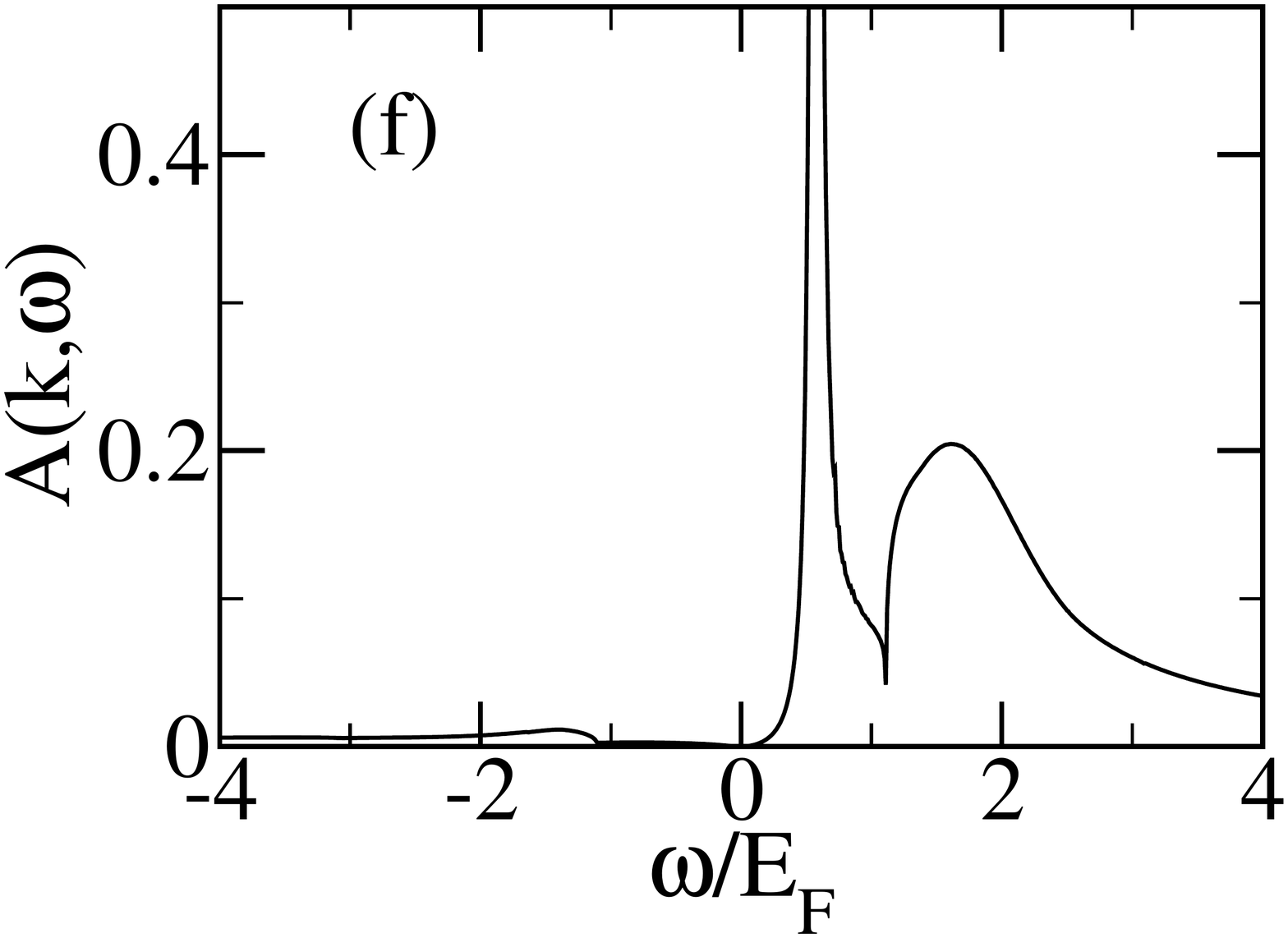}
\caption{(Color online) Left column: The real (thick black line) and imaginary part (dashed blue line) of self-energy for (a) $k=0.75k_F$, (c) $k=k_F$ and (e) $k=1.25k_F$. The absolute value of the imaginary part is plotted in these graphs. Right column: Single particle spectral function for the same wave vectors, (b) $k=0.75k_F$, (d) $k=k_F$ and (f) $k=1.25k_F$. The system is characterized by $r_s=3$. The Dirac point is at $\omega=-E_F$.}
\label{fig:spfn}
\end{figure}

The non-interacting Green function for the system is given by
$G_0(s,\kk,\omega)=(\omega-\xi_{s\kk})^{-1}$, where
$\xi_{s\kk}=E_{s\kk}-E_F$ with $E_F$ being the Fermi energy. The
corresponding spectral function consists of delta function peaks at
the band dispersion energies and is given by
$A_0(s,\kk,\omega)=-(1/\pi) Im
G_0(s,\kk,\omega)=\delta(\omega-\xi_{s\kk})$. Electron-electron
interactions cause scattering of the non-interacting excitations
resulting in the broadening of the delta function peaks and shifting
of spectral weight. This is incorporated through the modification of
the interacting Green function by a self energy term,
$G^{-1}(s,\kk,\omega)=G_0^{-1}(s,\kk,\omega)-\Sigma_s(\kk,\omega)$ and
the corresponding spectral function is given by
\beq
A(s,\kk,\omega)=-\frac{1}{\pi}\frac{ \Sigma_s^{''}(\kk,\omega)}{[\omega-\xi_{s\kk}-  \Sigma_s^{'} (\kk,\omega)]^2+[\Sigma_s^{''}(\kk,\omega)]^2}
\eeq
where $\Sigma_s(\kk,\omega+i0^+)=\Sigma_s^{'}(\kk,\omega)+ i
\Sigma_s^{''}(\kk,\omega)$.  Thus the
imaginary part of self-energy incorporates the broadening of the
spectral function peak while the real part determines the shift in
the energy dispersion, especially near the Fermi surface. 


We calculate the self-energy and hence the spectral
function of a doped BLG system within the dynamical RPA at zero temperature. Using Matsubara frequencies, the self-energy
can be written as
%
\bqa
\Sigma(s,\kk,i\omega)&=&-\frac{g}{\beta}\sum_{s',i\omega'}\int d^2 \qq \frac{V_c(\qq)}{\ep(\qq,i\omega')}\\
\no & &G_0(s',\kk-\qq,i\omega-i\omega')F_{ss'}(\kk,\qq)
\label{eqn:sigma_mat}
\eqa
where $i\omega$ and $i\omega'$ are respectively fermionic and bosonic
Matsubara frequencies. Here $V_c(q)=2\pi e^2/(\kappa q)$ is the 2D
bare Coulomb potential ($\kappa$ being the background dielectric
constant). Using $E_F$ as units of energy and $k_F$ as units of
momentum, the strength of the interaction can be written in terms of
the dimensionless coupling parameter $r_s=e^2gm/(\kappa k_F)$. Unlike SLG (but similar to 2DEG),
$r_s$ in BLG can be tuned by changing the density of carriers
($r_s\sim n^{-1/2}$) and we will present our results as a function of
$r_s$ rather than the density. The dielectric function
$\ep(\qq,\omega)$, which incorporates the effects of dynamic
screening, was calculated analytically in
Ref.~\onlinecite{sensarma:blg}. Finally, the chirality of the bands
is encoded in the overlap vertex factor $F_{ss'}(\kk,\qq)=(1+ss'\cos
2\theta)/2$, where $\theta$ is the angle between $\kk$ and $\kk+\qq$.
\begin{figure}
\includegraphics[width=0.23\textwidth]{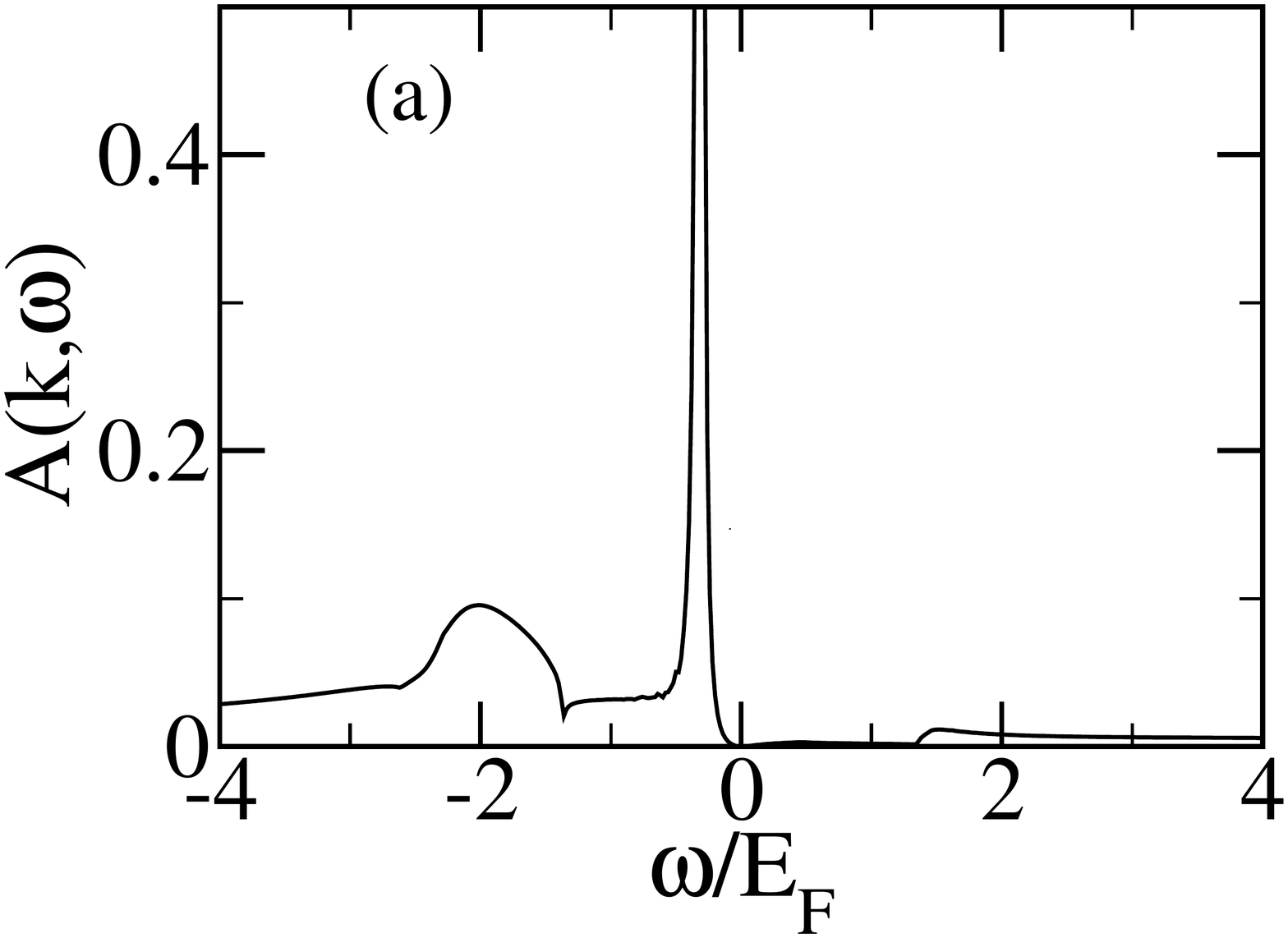}
\includegraphics[width=0.23\textwidth]{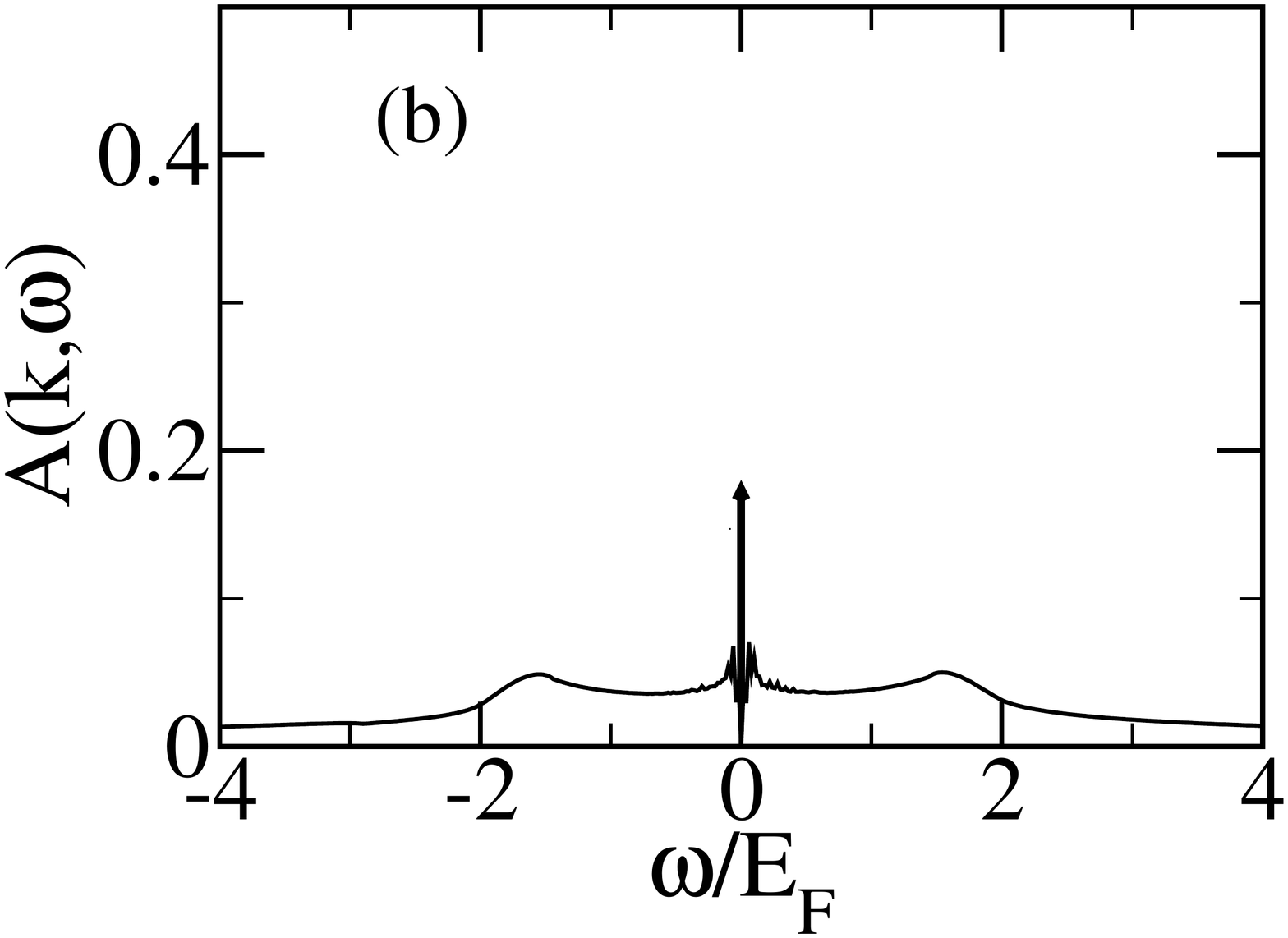}
\includegraphics[width=0.23\textwidth]{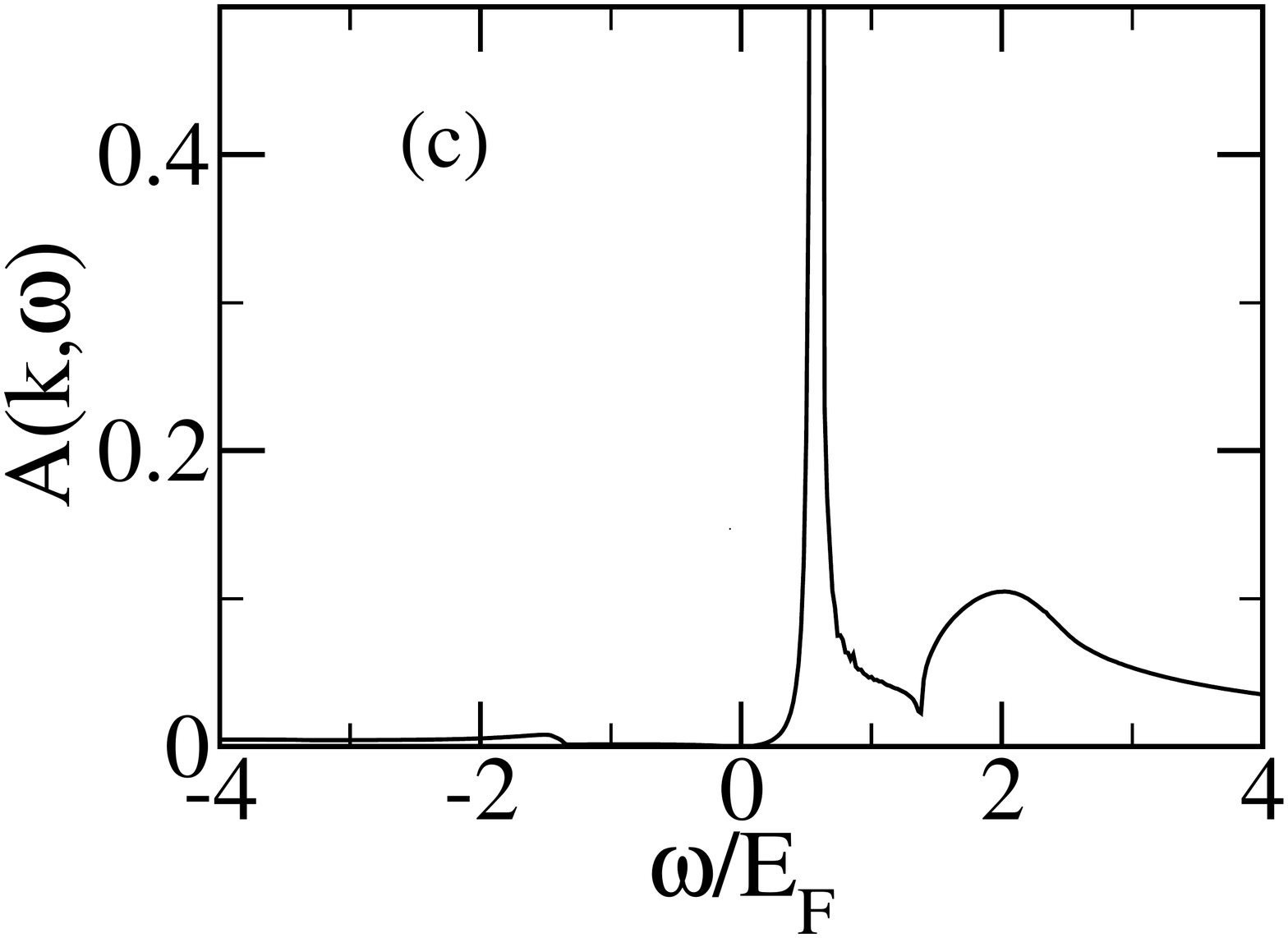}
\caption{ Single particle spectral function for the wave vectors, (a) $k=0.75k_F$, (b) $k=k_F$ and (c) $k=1.25k_F$ for a system characterized by $r_s=7$. The low energy quasiparticle peaks are accompanied by plasmaron features away from $k=k_F$.}
\label{fig:spfn1}
\end{figure}

Although it is customary to separate the self-energy into an exchange
and a correlation contribution, we prefer to work with the
full expression since the exchange contribution, by itself, is
divergent due to presence of the filled Fermi sea in the valence
band. The total contribution, however, is free of this divergence,
showing that the use of the dynamically screened interaction is crucial in calculating the single particle spectral properties of BLG.
It is important to emphasize that our dynamical RPA self-energy
calculation is theoretically well-defined and does not involve any
arbitrary infrared or ultraviolet regularization through arbitrary
momentum cut-offs.

{\it Spectral function}: With no loss of generality, we assume the BLG
chemical potential to be in the conduction band. Since we are
interested in the low energy properties around the Fermi surface, we
will focus on $A(+,\kk,\omega)$ and drop the index $s$ from now on.
In the left hand column of Fig.~\ref{fig:spfn} we plot the calculated
real and imaginary parts of the self-energy for three wavevectors, (a)
$k=0.75k_F$ (b)$k=k_F$ and (c) $k=1.25k_F$, in a system characterized
by $r_s=3$, corresponding to carrier density $n= 3\times
10^{12}cm^{-2}$ for $SiO_2$ substrate. The self-energy has two distinct
observable structures, coming from scattering off plasmon collective
modes (plasmarons) and the continuum of particle hole excitations. In
the imaginary part of the self-energy, the plasmon contribution leads
to distinct logarithmic singularities away from $k=k_F$. These
singularities are accompanied by jumps in the real part of the
self-energy. The continuum contribution leads to an increasing
background and dominates the self-energy at large values of
$\omega$. In the right hand column of Fig.~\ref{fig:spfn}, we plot the
corresponding BLG spectral function. We also show the spectral
function for a larger value of $r_s=7$ (lower density of $n= 6
\times 10^{11}cm^{-2}$) in Fig.~\ref{fig:spfn1}. The spectral
functions at these two values of $r_s$ show qualitatively similar
features, although for the larger value of $r_s$, the incoherent
background part of the spectral weight is spread over a larger
energy. The prominent feature of the spectral function, manifesting
its Fermi liquid characteristics, consists of a quasiparticle peak
which disperses across the Fermi energy as the wave-vector goes across
the Fermi surface. The quasiparticle peak is broadened for wavevectors
away from $k=k_F$. At $k=k_F$, the spectral function has a delta
function peak at the Fermi energy with a weight $Z\sim 0.4$ for
$r_s=3$ and $Z\sim 0.28$ for $r_s=7$. This is the classic Fermi liquid
behavior of an interacting electron system. Away from the Fermi
wavevector, the spectral function consists of a double peaked
structure. The second peak, arising from the dressing of the
quasiparticles by the plasmon modes, is often referred to in the
literature as a plasmaron. In SLG systems, these satellite bands have
already been observed in ARPES experiments~\cite{ARPESref}, and our
current work shows that both the quasiparticle peak and the plasmaron
should be manifestly observable in BLG as well.

\begin{figure}[t!]
\includegraphics[width=0.38\textwidth]{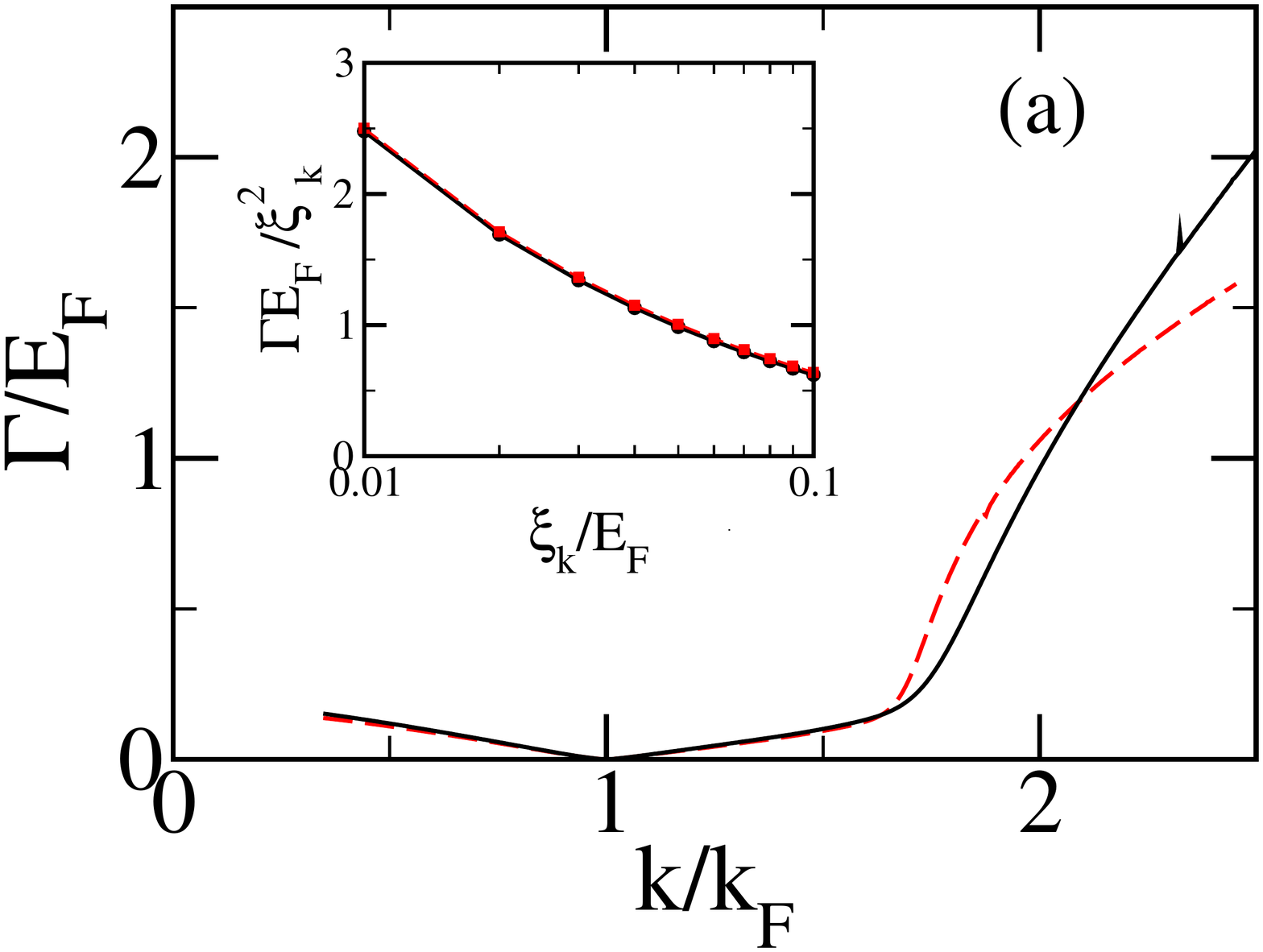}
\includegraphics[width=0.23\textwidth]{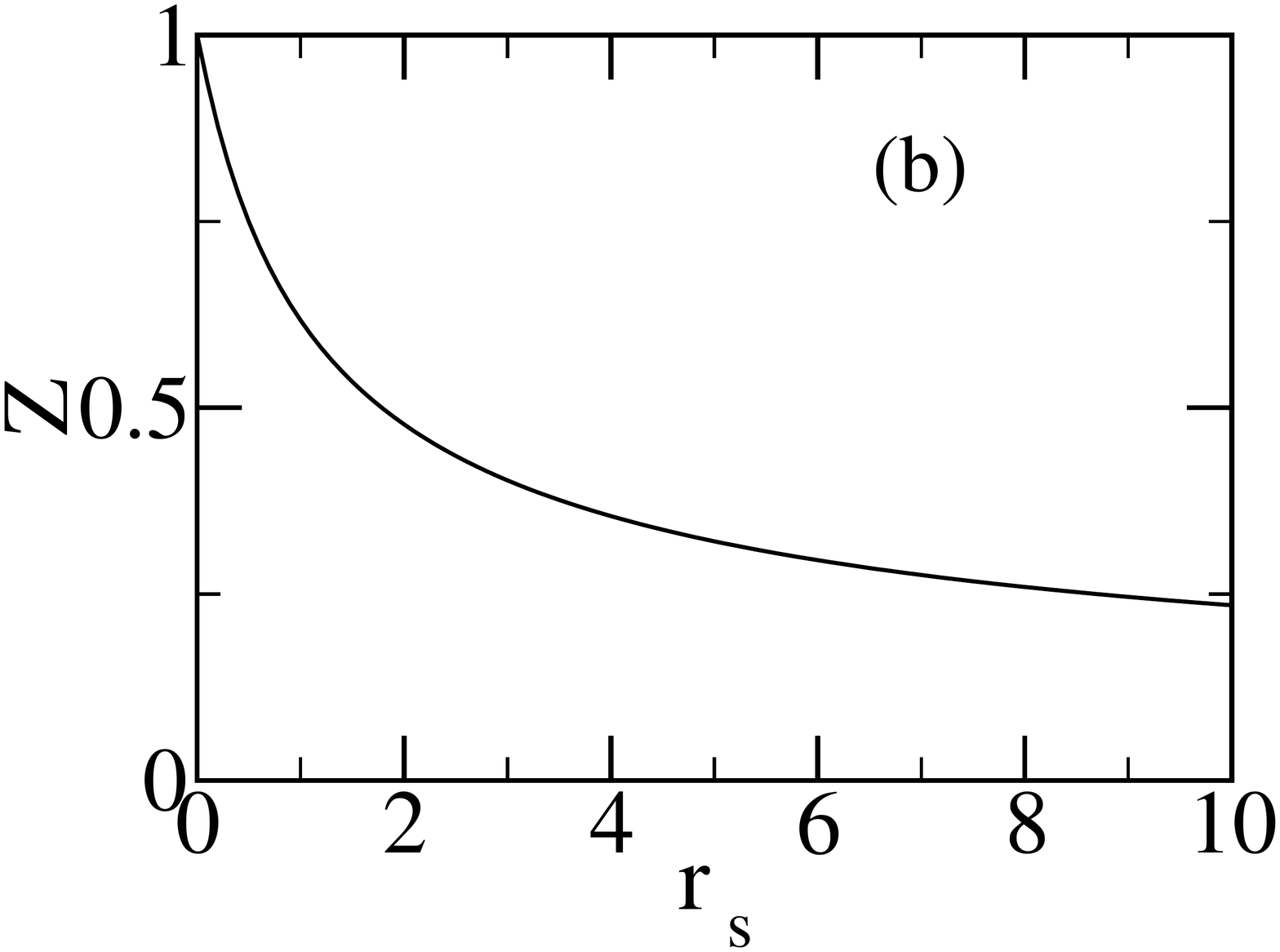}
\includegraphics[width=0.23\textwidth]{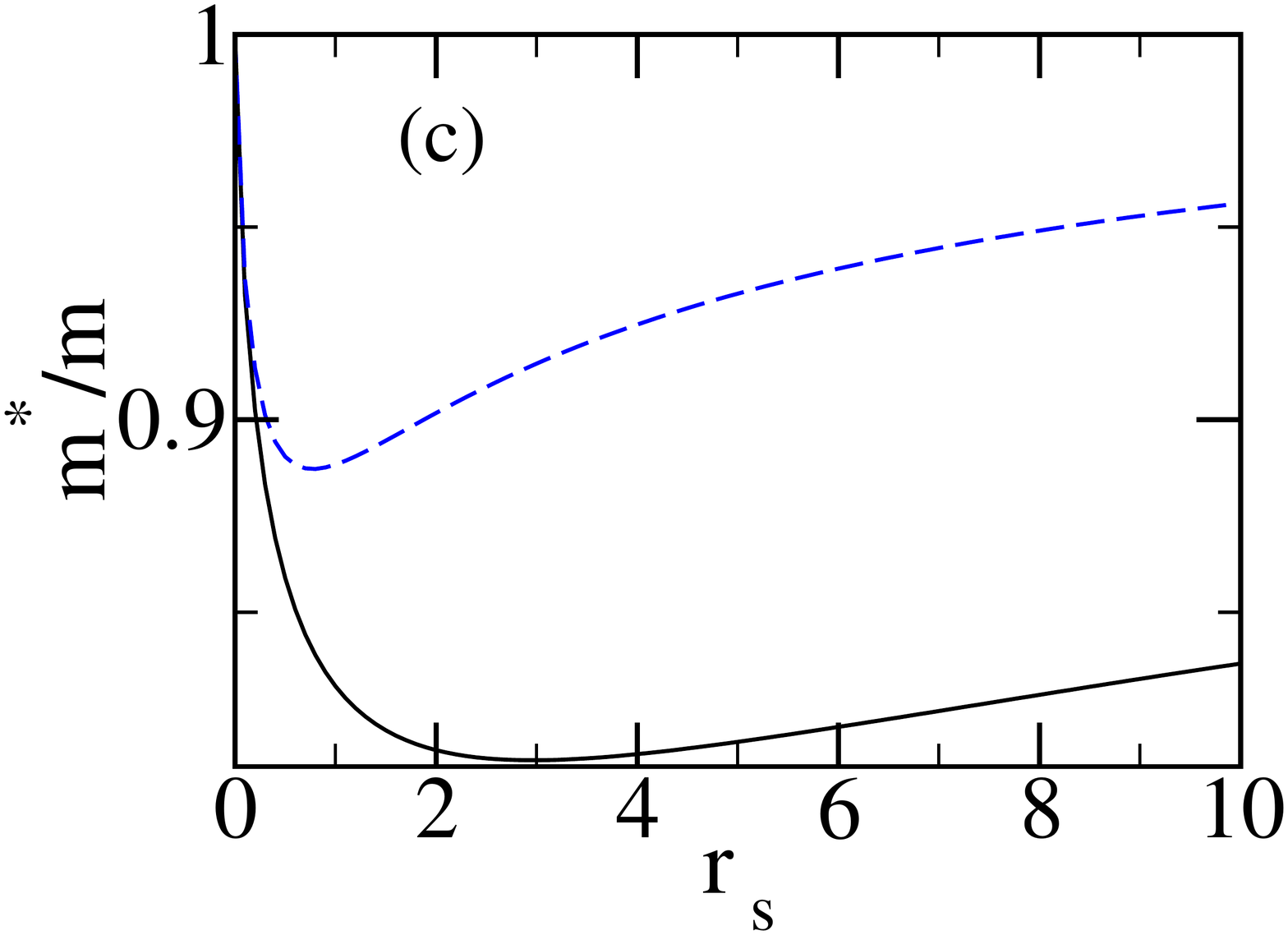}
\caption{(Color online): (a) Width of the quasiparticle peaks, $\Gamma_\kk= Im \Sigma(\kk,\xi_\kk)$, as a function of $k/k_F$ for two different values of $r_s$. The red dashed ine is for $r_s=3$ and the thick black line is for $r_s=7$. Inset: The low energy part of $\Gamma_\kk/\xi_\kk^2$ as a function of $\xi_\kk$ for $r_s=3$ and $r_s=7$. The x-axis is logarithmic and the linear plot shows the $\omega^2\ln \omega$ scaling of the quasiparticle width. (b) The quasiparticle weight, $Z$, as a function of $r_s$. (b) The effective mass $m^*$ as a function of $r_s$. In (b) the thick black line is obtained from the on-shell approximation whereas the blue dashed line is the self-consistent solution.  
}
\label{Fig:flpar}
\end{figure}

Landau's great insight in to the theory of interacting Fermi systems
was that due to phase space constraints arising from Pauli blocking,
the imaginary part of the self energy vanishes rapidly as one
approaches the Fermi surface and weakly interacting quasiparticles
form the low energy elementary excitations of the system. In a 2DEG,
at low energies, the width of the quasiparticle peak $\Gamma_\kk= Im
\Sigma(\kk,\xi_\kk)\sim (\xi_\kk^2/E_F)\ln (\xi_\kk/E_F)$. This
ensures that the spectral function has a delta function peak at the
Fermi surface with a finite weight $Z < 1$. For BLG, we analytically
find that the leading order quasiparticle scattering rate follows the
same asymptotic form, as shown in Fig.~\ref{Fig:flpar}(a). In fact
this can be analytically shown to be true for arbitrarily large values
of $r_s$. Thus the system remains a Fermi liquid for arbitrarily small
densities. Several authors~\cite{nfl} have recently shown that the
strictly undoped zero density neutral graphene is a non Fermi liquid
with a vanishing $Z$ factor and $Im \Sigma(k,\omega)\sim
\omega$. However this non Fermi liquid behaviour relies on two
peculiar features of undoped BLG : (a) a quadratic dispersion around a
single Fermi point and (b) the interband scattering contribution
persisting upto $\omega=0$. Both these conditions are violated at any
finite carrier density: the dispersion becomes linear around the Fermi
surface and the interband scattering contribution is cut-off at a
scale of $E_F$. The low energy quasiparticle width is in fact
dominated by forward scattering from intraband particle-hole continuum
and follows the usual Fermi liquid form. In the renormalization group language, the
chemical potential is a relevant perturbation (even at the tree level)
and thus the non Fermi liquid fixed point is unstable to arbitrarily
small densities. We therefore conclude that the BLG is always a Fermi
liquid.

Thus, for the doped BLG near the Fermi surface, the
spectral function can be written as
\beq
A(\kk,\omega)=Z_k\delta(\omega-v_F k+\mu)+A^{inc}(\kk,\omega)
\eeq
where $\mu$ is the chemical potential of the system, $v_F$ is the
slope of the dispersion, $Z$ is the quasiparticle spectral weight
and $A^{inc}$ the non-singular incoherent part of the spectral
function. The effective mass of the quasiparticles, which enters into
low temperature thermodynamic properties like specific heat, is
defined as $m^*=k_F/v_F$.  The self-energy at low energies thus
contains important information about the quasiparticle dispersion,
broadening and spectral weight.

{\it Quasiparticle spectral weight}: As mentioned before, phase space
restrictions associated with the fermionic Pauli principle lead to
sharply defined quasiparticles near the Fermi surface in the
interacting system producing the one-to-one Fermi liquid
correspondence with the non-interacting Fermi gas. In BLG, the
imaginary part of the self-energy at $k=k_F$ vanishes as $\omega^2\ln
\omega$ as one approaches the Fermi energy and exactly at the Fermi
surface, the spectral function has a delta function peak with a weight $Z \leq 1$. This leads to a discontinuity in the
momentum distribution at the Fermi surface (at $T=0$) with the
magnitude of the discontinuity given by $Z$. In terms of the self
energy the quasiparticle spectral weight is given by
\beq
 Z^{-1}= 1-\frac{\partial \Sigma^{'}(k_F,\omega)}{\partial \omega}\vert_{k_F,E_F}.
\eeq 
%
In Fig.~\ref{Fig:flpar}(a), we plot the BLG quasiparticle
weight as a function of $r_s$. The quasiparticle
weight decreases monotonically with increasing $r_s$ (decreasing
density), as stronger interactions shift spectral
weight away from the quasiparticle pole. For very large $r_s$, $Z$ eventually slowly approaches zero asymptotically with the system remaining a Fermi liquid for all $r_s$ except the zero density infinite $r_s$ singular point.

 The energy dispersion of the
quasiparticle excitations can be extracted either from the
self-consistent solution of Dyson's equation (often called the off-shell approximation)
\beq \omega_k-\xi_\kk-\Sigma^{'}(\kk,\omega_k)=0 \eeq
or from the on-shell approximation (OSA)
\beq
\omega_k-\xi_\kk-\Sigma^{'}(\kk,\xi_k)=0
\eeq
It is clear that if the self-energy is calculated exactly, the Dyson
equation must be solved self-consistently.  However the RPA involves a
single-loop leading-order approximation in the dynamically screened
Coulomb interaction and as such, the
dispersion should be calculated within the OSA to get consistent
results within RPA~\cite{Rice}.
However, since the use of the off-shell approximation is found in
various places in the literature, we present our results both within the
OSA and the off-shell Dyson equation approximations. We find that the
interaction effects for
the off-shell approximation are smaller than those for the OSA. We believe that experimental measurements of effective mass and/or the $Z$-factor compared with our theory could shed light on the important question of the relative validity of on-shell versus off-shell approximation.

{\it The effective mass ($m^*$)}: The slope
of the quasiparticle dispersion around the Fermi surface determines
the Fermi velocity $v_F$ which can be written in terms of the
effective mass of the Landau quasiparticles as $v_F=k_F/m^\ast$. The
self consistent solution to the Dyson equation yields
\beq
\frac{m^*}{m_0}=\frac{Z^{-1}}{1+\frac{m_0}{k_F}\frac{\partial \Sigma^{'}(k,\omega)}{\partial k}\vert_{k_F,E_F}}
\eeq
where $Z$ is the quasiparticle weight. The on-shell approximation gives
\beq
\frac{m^*}{m_0}=\frac{1}{1+\frac{m_0}{k_F}\frac{\partial \Sigma^{'}(k,\omega)}{\partial k}\vert_{k_F,E_F}+\frac{\partial \Sigma^{'}(k,\omega)}{\partial \omega}\vert_{k_F,E_F}}
\eeq
The effective mass obtained from both calculations is plotted as a
function of $r_s$ in Fig.~\ref{Fig:flpar}(b). We find that the
effective mass at first decreases with increasing $r_s$ and then exhibits a weak
nonmonotonicity for larger $r_s$, where it tends to increase a bit, with
the renormalized mass being always less than the bare band mass even
for $r_s$ as large as $10$.  This is very different from 2DEG and 3DEG
systems where the effective mass renormalizes to values much larger
than the bare mass for $r_s$ larger than unity~\cite{Rice}.  This
prediction of our theory, which is somewhat counterintuitive (i.e. the
renormalized mass being less than the bare mass even for strong
interaction), is easily experimentally verifiable.  We note that the
BLG effective mass renormalization is small (only of the order of
$\sim 20\%$ or less) in sharp contrast to 2DEG or 3DEG where the mass
renormalization could be by a factor of $5$ or more at larger
$r_s$~\cite{Rice}.  The relatively small BLG mass renormalization
indicates a larger regime of validity for dynamical RPA in the chiral gapless 
BLG system compared with the standard 2DEG system.  The maximum mass
renormalization of $\sim 20 \%$ also shows that the system remains a
weakly interacting Fermi liquid upto a relatively large $r_s$.


{\it Conclusion}: We have calculated the effects of electron-electron
interaction on the single particle spectral function of bilayer
graphene within the random phase approximation. The spectral function
shows a clear satellite plasmaron peak in addition to the Landau
quasiparticle peak near the Fermi surface. Thus, it should be possible
to obtain information about collective plasmon excitations in the
system by looking for these satellite peaks in ARPES data. Focusing on
the low energy properties of the Landau quasiparticles, we have
calculated the quasiparticle weight and the effective mass of the
Landau quasiparticles as a function of $r_s$. We show that inclusion
of dynamic screening is crucial in obtaining these results. We also
show that the presence of the valence band leads to relatively small
renormalization in the low energy spectral properties, and thus the
system remains a weakly interacting Fermi liquid upto relatively large
values of $r_s$ or relatively low densities. In addition to obtaining
the quasiparticle spectral function, the effective mass
renormalization, the inelastic scattering rate, and the quasiparticle
renormalization factor, we have established the theoretical
principle that doped BLG is always a Fermi liquid in contrast to
undoped BLG.

This work is supported by ONR-MURI and NRI-SWAN.

\end{document}